\documentstyle[preprint,aps,tighten,floats,epsf]{revtex} 
\preprint{T97/044}
\begin{document} 
\draft
\title{PERCOLATION TRANSITION IN THE RANDOM ANTIFERROMAGNETIC 
SPIN-1 CHAIN}
\author{C. Monthus, O. Golinelli and Th. Jolic\oe
ur\thanks{e-mail: monthus, golinel, thierry@spht.saclay.cea.fr}}
\address{Service de Physique Th\'eorique, CEA Saclay,\\
F91191 Gif-sur-Yvette, France} 
\date{\today} 
\maketitle
%%%%%%%%%%%%%%%%%%%%%%%%%%%%%%%%%%%%%%%%%%%%%%%%%%%%%%%%%%%%%%%%%
\begin{abstract} 
We give a physical description in terms of
percolation theory of the phase transition that occurs when the 
disorder increases in the random
antiferromagnetic spin-1 chain between a gapless phase with 
topological order and a random singlet phase.
We study the statistical
properties of the percolation clusters by numerical simulations, and we compute exact exponents characterizing the 
transition by a real-space renormalization group 
calculation. 
\end{abstract}
%%%%%%%%%%%%%%%%%%%%%%%%%%%%%%%%%%%%%%%%%%%%%%%%%%%%%%%%%%%%%%%%%
Antiferromagnetic quantum spin chains are
known to present vastly different physical properties according to
whether the spin value is integer or half-integer\cite{FDM}. In the
case of integer spin, there is a (Haldane) gap for magnetic
excitations and short-range magnetic order. In the case of
half-integer spin, there is quasi-long range spin order with
algebraically decaying spin correlations and no gap. An appealing
physical picture of the integer-spin case is given by the so-called
valence-bond-solid (VBS) wavefunctions\cite{AKLT}. With these
approximate ground states, it is easy to see that there is a hidden
long-range order that is non-local in terms of the true
spins-1\cite{SOP}. These properties are well established for the pure
systems but it is obviously important to understand what happens in
the disordered case. Bond disorder in a quantum spin chain may be
realized experimentally by chemical substitution of the ligand that
induces superexchange between the spins, for example. This leads to a Heisenberg Hamiltonian: 
\begin{equation} 
H=\sum_i J_i \,\, {\vec S}_i
\cdot {\vec S}_{i+1}, 
\label{ham} 
\end{equation} 
where 
$\vec S_i$ are quantum spin operators and the couplings $J_i$ are
quenched random variables. In the spin-1/2 case, it is known\cite{MDH,Fisher}
that for any strength of the disorder the system
goes to the so-called random singlet phase\cite{BhattLee} in which the spins are
paired into singlets over arbitrarily large distances.
The effect of randomness on other quantum spin chains has also
been investigated recently\cite{West,Ha,Hb,Boe,Hida}.

In the spin-1 case, it has been shown\cite{HY,PhD} that this random
singlet phase appears only when the randomness is strong enough. 
These results have been obtained by an asymptotically exact
real-space renormalization group (RG) study on an effective 
spin-{1/2} model. For weak disorder, there is a phase which is 
gapless but sustains hidden long-range order. This is analogous to 
gapless superconductivity that occurs when doping a superconductor 
with magnetic impurities; here the role of the superconducting 
condensate is played by the hidden order. 

In this Letter, we use a real-space renormalization procedure that
leads to an appealing percolation picture of the phase transition
that occurs in the disordered antiferromagnetic spin-1 chain. We
formulate the hidden long-range order in terms of a macroscopic
VBS-cluster. This allows us to define several critical exponents to
describe this phase transition. We study them by direct numerical
renormalization and compare to the exact
values that we obtain for the effective model of Ref.\,\cite{HY} 
which is in the same universality class.

An efficient technique to study random spin chains is the real-space 
renormalization group\cite{MDH,Fisher}. The basic idea is to 
integrate out the strongest bond of the chain and to compute in
perturbation theory the corresponding renormalization of the 
coupling strengths distribution. This can be done iteratively
and, in the case of the S=1/2 chain, leads to a universal fixed 
point distribution which is extremely broad, the so-called random singlet phase\cite{MDH}. However, when generalized to higher spin values, this method is blind to the difference between integer and half-integer spins. To capture the physics of the Haldane phase of the random spin-1 chain, we need to
introduce an extended set of spin chains described by the
Hamiltonian (\ref{ham}), where $\vec S_i$ is a spin-operator of size
$s_i=1/2$ or $s_i=1$, and where the couplings $\{J_i\}$ can be either ferromagnetic (F) or antiferromagnetic (F), but have to satisfy the following constraint :
for any chain segment $\{i,j\}$, the classical magnetization must 
satisfy $\vert m_{i,j} \vert \leq 1$. 

This condition for $j=i+1$ implies that there are four types
of bonds : 
i) F bond between two spins S=1/2,
ii) AF bond between two spins S=1/2,
iii) AF bond between one spin S=1 and one spin S=1/2, 
iv) AF bond between two spins S=1. 
The decimation procedure is the following : to each
bond $(i)$, we associate the energy difference $\Delta_i$ between 
the highest state and the lowest state of the Hamiltonian $J_i \vec
S_i \cdot \vec S_{i+1}$. We pick up the bond $\left( \vec S_{i}, 
\vec S_{i+1} , J_{i} \right)$ corresponding to the strongest 
$\Delta_i$ of the chain. To define the renormalization rule for this bond, we
divide the four-spin Hamiltonian into 
$H_i= h_0+ h_1$ 
where 
$ h_0=J_{i} \vec S_{i} \cdot\vec S_{i+1} $ 
and 
$ h_1= J_{i-1}\vec S_{i-1} \cdot \vec S_{i}+ J_{i+1} \vec S_{i+1} 
\cdot \vec S_{i+2} $.
We treat $h_1$ as a perturbation of $h_0$ to find the effective
Hamiltonian replacing $H_i$ when the highest energy
state of $h_0$ is removed. This leads to the four following
renormalizations rules:

\bigskip
\bigskip
\hbox to 450pt {(1)\hfill
\vbox{\hsize=10pt \centerline{$s_0$}\par \centerline{$\bullet$}} 
\kern -7pt \raise 2pt
\vtop{\hsize=45pt \centerline{\hrulefill} \par \centerline{$J_0$}} 
\kern -7pt
\vbox{\hsize=10pt \centerline{$s_1={1\over 2}$}\par 
\centerline{$\bullet$}} \kern -7pt \raise 2pt
\vtop{\hsize=45pt \centerline{\hrulefill} \par \centerline{$J_1<0$}} 
\kern -7pt
\vbox{\hsize=10pt \centerline{$s_2={1\over 2}$}\par 
\centerline{$\bullet$}} \kern -7pt \raise 2pt
\vtop{\hsize=45pt \centerline{\hrulefill} \par \centerline{$J_2$}} 
\kern -7pt
\vbox{\hsize=10pt \centerline{$s_3$}\par \centerline{$\bullet$}} 
\hfill$\longrightarrow$\hfill
\vbox{\hsize=10pt \centerline{$s_0$}\par \centerline{$\bullet$}} 
\kern -7pt \raise 2pt
\vtop{\hsize=67pt \centerline{\hrulefill} \par 
\centerline{$J^\prime_0={J_0\over 2}$}} \kern -7pt
\vbox{\hsize=10pt \centerline{$s^\prime_1=1$}\par 
\centerline{$\bullet$}} \kern -7pt \raise 2pt
\vtop{\hsize=67pt \centerline{\hrulefill} \par 
\centerline{$J^\prime_1={J_2\over 2}$}} \kern -7pt
\vbox{\hsize=10pt \centerline{$s_3$}\par \centerline{$\bullet$}} 
\hfill}
\bigskip
\bigskip
\hbox to 450pt {(2)\hfill
\vbox{\hsize=10pt \centerline{$s_0$}\par \centerline{$\bullet$}} 
\kern -7pt \raise 2pt
\vtop{\hsize=45pt \centerline{\hrulefill} \par \centerline{$J_0$}} 
\kern -7pt
\vbox{\hsize=10pt \centerline{$s_1={1\over 2}$}\par 
\centerline{$\bullet$}} \kern -7pt \raise 2pt
\vtop{\hsize=45pt \centerline{\hrulefill} \par \centerline{$J_1>0$}} 
\kern -7pt
\vbox{\hsize=10pt \centerline{$s_2={1\over 2}$}\par 
\centerline{$\bullet$}} \kern -7pt \raise 2pt
\vtop{\hsize=45pt \centerline{\hrulefill} \par \centerline{$J_2$}} 
\kern -7pt
\vbox{\hsize=10pt \centerline{$s_3$}\par \centerline{$\bullet$}} 
\hfill$\longrightarrow$\hfill
\vbox{\hsize=10pt \centerline{$s_0$}\par \centerline{$\bullet$}} 
\kern -7pt \raise 2pt
\vtop{\hsize=135pt \centerline{\hrulefill} \par 
\centerline{$J^\prime_0={J_0\,J_2\over 2\,J_1}$}} \kern -7pt
\vbox{\hsize=10pt \centerline{$s_3$}\par \centerline{$\bullet$}} 
\hfill}
\bigskip
\bigskip
\hbox to 450pt {(3)\hfill
\vbox{\hsize=10pt \centerline{$s_0$}\par \centerline{$\bullet$}} 
\kern -7pt \raise 2pt
\vtop{\hsize=45pt \centerline{\hrulefill} \par \centerline{$J_0$}} 
\kern -7pt
\vbox{\hsize=10pt \centerline{$s_1=1$}\par \centerline{$\bullet$}} 
\kern -7pt \raise 2pt
\vtop{\hsize=45pt \centerline{\hrulefill} \par \centerline{$J_1>0$}} 
\kern -7pt
\vbox{\hsize=10pt \centerline{$s_2={1\over 2}$}\par 
\centerline{$\bullet$}} \kern -7pt \raise 2pt
\vtop{\hsize=45pt \centerline{\hrulefill} \par \centerline{$J_2$}} 
\kern -7pt
\vbox{\hsize=10pt \centerline{$s_3$}\par \centerline{$\bullet$}} 
\hfill$\longrightarrow$\hfill
\vbox{\hsize=10pt \centerline{$s_0$}\par \centerline{$\bullet$}} 
\kern -7pt \raise 2pt
\vtop{\hsize=45pt \centerline{\hrulefill} \par 
\centerline{$J^\prime_0={4\,J_0\over 3}$}} \kern -7pt
\vbox{\hsize=10pt \centerline{$s^\prime_1={1\over 2}$}\par 
\centerline{$\bullet$}} \kern -7pt \raise 2pt
\vtop{\hsize=90pt \centerline{\hrulefill} \par 
\centerline{$J^\prime_1=-{J_2\over 3}$}} \kern -7pt
\vbox{\hsize=10pt \centerline{$s_3$}\par \centerline{$\bullet$}} 
\hfill}
\bigskip
\bigskip
\hbox to 450pt {(4)\hfill
\vbox{\hsize=10pt \centerline{$s_0$}\par \centerline{$\bullet$}} 
\kern -7pt \raise 2pt
\vtop{\hsize=45pt \centerline{\hrulefill} \par \centerline{$J_0$}} 
\kern -7pt
\vbox{\hsize=10pt \centerline{$s_1=1$}\par \centerline{$\bullet$}} 
\kern -7pt \raise 2pt
\vtop{\hsize=45pt \centerline{\hrulefill} \par \centerline{$J_1>0$}} 
\kern -7pt
\vbox{\hsize=10pt \centerline{$s_2=1$}\par \centerline{$\bullet$}} 
\kern -7pt \raise 2pt
\vtop{\hsize=45pt \centerline{\hrulefill} \par \centerline{$J_2$}} 
\kern -7pt
\vbox{\hsize=10pt \centerline{$s_3$}\par \centerline{$\bullet$}} 
\hfill$\longrightarrow$\hfill
\vbox{\hsize=10pt \centerline{$s_0$}\par \centerline{$\bullet$}} 
\kern -7pt \raise 2pt
\vtop{\hsize=45pt \centerline{\hrulefill} \par 
\centerline{$J^\prime_0=J_0$}} \kern -7pt
\vbox{\hsize=10pt \centerline{$s^\prime_1={1\over 2}$}\par 
\centerline{$\bullet$}} \kern -7pt \raise 2pt
\vtop{\hsize=45pt \centerline{\hrulefill} \par 
\centerline{$J^\prime_1=J_1$}} \kern -7pt\vbox{\hsize=10pt 
\centerline{$s^\prime_2={1\over 2}$}\par \centerline{$\bullet$}} 
\kern -7pt \raise 2pt
\vtop{\hsize=45pt \centerline{\hrulefill} \par 
\centerline{$J^\prime_2=J_2$}} \kern -7pt
\vbox{\hsize=10pt \centerline{$s_3$}\par \centerline{$\bullet$}} 
\hfill}
\bigskip
\bigskip

These rules are a slight modification of those proposed by 
Hyman\cite{PhD}.
Our renormalization procedure is entirely consistent from the point
of view of the progressive elimination of the highest energy degrees
of freedom : all the energy scales of the new bonds are
always smaller than the energy scale of the old bond. It is also 
easy to check that this 
procedure is ``closed" inside the enlarged set of spin-chains defined
above. In particular, spins higher than $1$ cannot appear in
this scheme\cite{long}.

If we represent each spin-1 as the symmetrization of two spin-1/2, the rules 2, 3, and 4 for AF bonds may
then be interpreted as the formation of a singlet between
two constitutive spin-1/2. The rule 1 for a
F bond between two spin-1/2 corresponds to their
symmetrization. As a consequence, at
the end of the renormalization procedure, when there are no free
spins left, the chain decomposes itself into a set of disconnected
clusters that have the structure of a VBS-state (see Figure 1).
\begin{figure}
\centering\leavevmode
\epsfxsize=10cm
\epsfbox{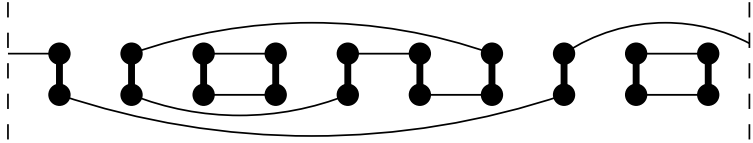}

\caption{The ground state of the random chain obtained from the
real-space renormalization. Each spin S=1 is written as two symmetrized
spins S=1/2: the black dots related by a vertical thick line. Spins are
grouped in clusters related by singlet bonds that are the thin lines.
There are two clusters of size 2, one of size 4, and one larger cluster
that extends beyond the limits of the figure.}

\end{figure}
The topological order can be probed by use of the string order
parameter\cite{SOP} defined as: 
\begin{equation} 
t_{ij}= - \langle\psi_o
\big \vert S_i^z \exp\left[i \pi \sum_{i<k<j} S_k^z \right]
S_{j}^z\big\vert \psi_0\rangle .
\end{equation} 
For the pure VBS state on a
finite ring of $n$ spins S=1, one has $t_{ij}=4/9$ (up to 
corrections
of O(3$^{-n})$ that we neglect in the following). Now for the
disordered system, one has 
$t_{ij}=4/9$ if the two sites do belong to the same
cluster and $t_{ij}=0$ if they don't.
We thus consider, for a given realization of the disorder on
a chain of N spins, the spatial average of the string order parameter
$\sum_{i,j} t_{ij} /N^2$ which is equal to $4/9\times T$ where:
\begin{equation} 
T \equiv \sum_c {n_c^2\over N^2}={9\over 4} {1\over
N^2} \sum_{i,j} t_{ij} +O(1/N). 
\end{equation} 
Here c is the cluster
index and $n_c$ is the number of spins of the cluster, i.e. its size 
(which is not directly related to its spatial extent).
This quantity T is the probability that two randomly chosen spins belong to the same cluster. It is also
 the mean size of the cluster containing a
randomly chosen spin, divided by $N$. This order parameter can be
non-zero in the large-$N$ limit only if at least one VBS-cluster
contains a finite fraction of the spins of the chain.

We have simulated the renormalization procedure on spin-1 chains
containing $N$ sites with periodic boundary conditions. The initial
couplings $J_i$ are uniformly distributed in the interval $[1,1+d]$. 
The parameter $d$ measures the strength of the initial disorder. We
have numerically implemented the renormalization rules on 
$M$ initial independent samples, and averaged
quantities over these different realizations of the disorder.
Typically the averaging process has been performed on $N\times M
\approx 10^9$ spins. The plot of $T$ as a function of the disorder
for various sizes in Figure 2
\begin{figure}
\centering\leavevmode
\epsfxsize=12cm
\epsfbox{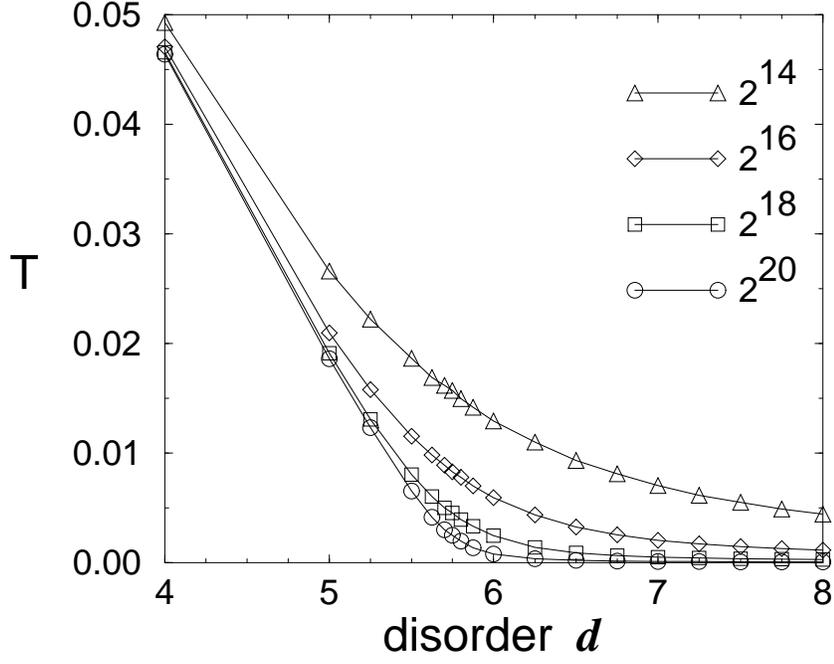}

\caption{String order parameter T as a function of the disorder d for
sizes $N=2^{14}-2^{20}$. The critical point is $d_c = 5.76(2)$}

\end{figure}
shows clearly that the phenomenology for the clusters is similar to the physics of percolation.
Indeed we observe a small-disorder phase in which there is exactly one macroscopic cluster (c=1) that contains a nonzero fraction 
$n_1 /N$
of the spins in the thermodynamic limit. There is also a distribution of finite clusters, i.e. non-diverging with N. This phase with $T\neq 0$ has long-range hidden order. 
There is a critical value $d_c$ of the disorder 
for which the diverging cluster disappears, i.e. T vanishes.
Then, for larger randomness $d>d_c$, there are only finite clusters
and $T=0$.

Following the
notations of percolation theory \cite{stauffer}, we denote by $\beta$
the exponent describing the vanishing of the fraction 
$n_1 /N$ of spins in the macroscopic cluster. The order parameter thus scales as $T \sim (d_c -
d)^{2\beta}$ for $d<d_c$. The finite size scaling study of Fig. 2
leads to the following  estimate: 
\begin{equation}
2\beta = 1.0(1). 
\label{nbeta} 
\end{equation} 
On Figure 3,
\begin{figure}
\centering\leavevmode
\epsfxsize=12cm
\epsfbox{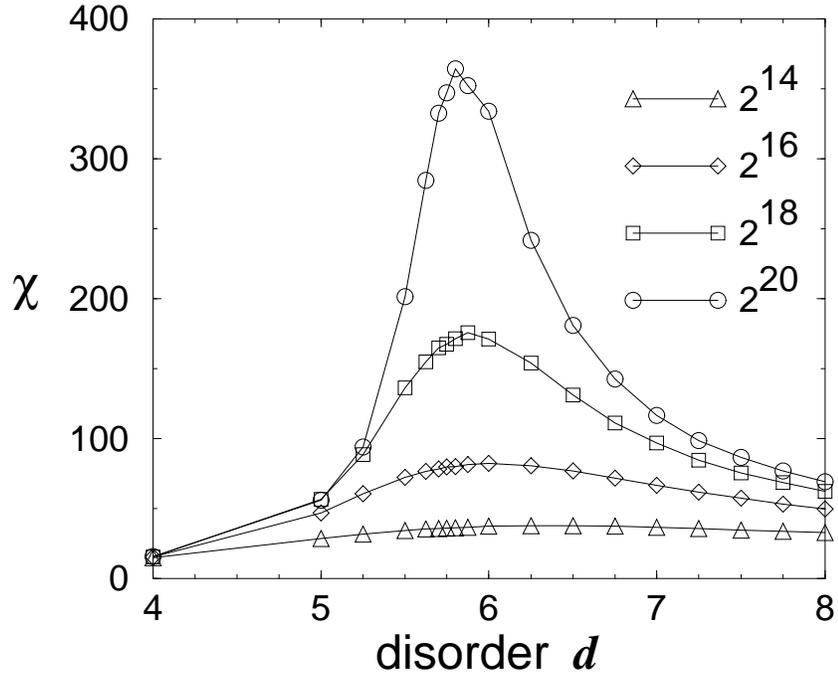}

\caption{Percolation susceptibility $\chi$, i.e. mean size of finite
clusters, as a function of disorder, for sizes $N=2^{14}-2^{20}$.}

\end{figure}
we have plotted the mean size of finite clusters, which 
plays the role of the percolation susceptibility: 
\begin{equation} 
\chi \equiv 
\sum_{c >1} {n_c^2 \over N} . 
\end{equation} 
In the sum, note that we omit the
biggest cluster (c=1). Denoting by $\gamma$ the exponent describing the
divergence of this quantity, $\chi \sim |d_c -d
|^{-\gamma}$, we have performed a finite size scaling study 
which leads to the value :
\begin{equation} \gamma =1.2(1). 
\label{numegamma}
\end{equation}

Let us call $m_d(s)$ the number of clusters of size $s$, divided
by N. Then standard scaling leads to the formula valid in the limit $N \to \infty$ \cite{stauffer} :
\begin{equation} 
m_d(s) \sim \
{ 1 \over {s^{\tau}}} \ \ f \left[ (d_c-d) s^{\sigma} \right]
\label{scaling} 
\end{equation} 
where $f(z)$ is a scaling
function. The exponent $\sigma$ defines the divergence of the 
characteristic cluster size as
 $\vert d_c-d \vert ^{- {1 / \sigma}}$ and the exponent $\tau$ characterizes the
algebraic decay of the distribution of cluster sizes at the critical
disorder $d_c$. These exponents can be related to $\beta$ and $\gamma$ as in percolation theory according to $\beta={{(\tau - 2)} / \sigma} $ and
$ \gamma={{(3-\tau )} / \sigma}$. In Fig. 4,
\begin{figure}
\centering\leavevmode
\epsfxsize=12cm
\epsfbox{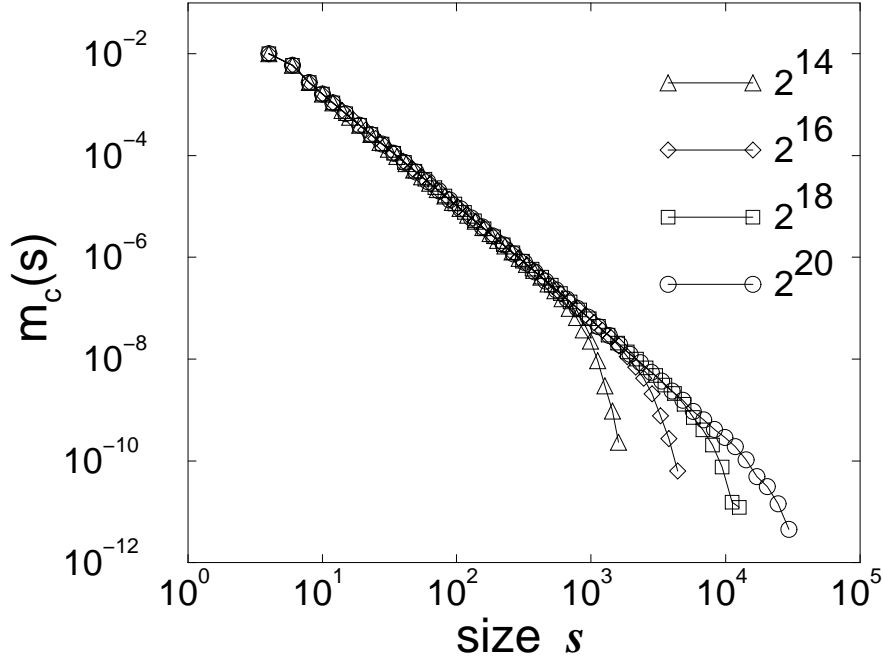}

\caption{Cluster size distribution $m_c (s)$  as a function of size s,
at criticality. The slope leads to $\tau =2.2(1)$.}

\end{figure}
we have plotted the
distribution of cluster sizes at criticality $d=d_c$.
Measurement of the slope leads to the following 
estimate of the exponent $\tau$: 
\begin{equation} 
\tau= 2.2(1).
\label{numetau} 
\end{equation}

We now show that the exponents that appear naturally in the percolation picture can be obtained exactly.
First, we  explain how the
VBS-clusters appear in the random dimerized
spin-{1/2} chain model of Hyman and Yang \cite{HY}, which is an
effective model for the random AF spin-1 chain. We
briefly recall the properties of the effective model that are
necessary for our purposes. In this model, all even bonds are
AF, whereas the odd bonds are either F
 or  AF. At any stage of their renormalization
procedure, the energy scale $\Omega$ is given by the strongest AF
bond of the system, so that the odd bonds separate into two groups :
group A contains all AF bonds and F bonds weaker than $\Omega$, while
group B contains all F bonds stronger than $\Omega$. In this
effective model, perfect VBS order means singlets
over all even bonds of the initial chain. Indeed, at the stable 
fixed
point of the Haldane phase found in \cite{HY} by solving the
renormalization flow equations, all 
odd
bonds are much weaker than even bonds so that only singlets over 
even
bonds are generated by decimation. It is  convenient to
introduce in their procedure an auxiliary variable
$\mu$ for each odd bond still surviving at scale $\Omega$. The
variable $\mu$ is by definition the number of singlets already made
over even bonds of the initial chain that are contained in this odd
bond. The evolution rules for this new variable are the
following : when an odd bond of variable ($\mu$) is decimated, a
finite cluster of size $(\mu+1)$ is terminated; when an even bond
surrounded by two odd bonds of group B of variables $\mu_1$ and
$\mu_2$ respectively, a finite cluster of size $(\mu_1+\mu_2+2)$
is terminated; when an even bond surrounded by two odd bonds of
group A, or surrounded by one odd bond of group A or one odd bond of
group B, with respective variables $\mu_1$ and $\mu_2$, the new odd
bond generated by this decimation inherits the variable
($\mu=\mu_1+\mu_2+1$). Following the method outlined by Fisher
\cite{Fisher}, we
find that, at the fixed point describing the transition of the effective model, the
auxiliary variable $\mu$ scales with $\Gamma=\ln\left({\Omega_0 /
\Omega}\right)$ (where $\Omega_0$ is the initial cut-off) as :
\begin{equation} 
\mu \propto \Gamma^{\varphi}
\qquad \hbox{with} \ \ \varphi=\sqrt 5 .
\end{equation} 
This has to be
compared with the scaling $l \propto \Gamma^3$ of the auxiliary
variable $l$ that counts the number of initial bonds in a surviving 
bond at scale $\Gamma$. 
Deviations from the critical point are driven by a relevant perturbation\cite{HY} that scales as $\Gamma^{\lambda_+}$ with 
$\lambda_+=({\sqrt{13} -1})/2$.
As a consequence, the exact exponent for the string topological order parameter (\ref{nbeta}) is\cite{active} :
\begin{equation} 
2\beta ={ {2(3-\varphi)} \over
{\lambda_+} } = { {4(3-\sqrt 5)} \over {\sqrt{13} -1} } = 
1.17278...,
\end{equation} 
 the
exponent of the percolation susceptibility is :
\begin{equation}
\gamma={{(2\varphi-3)} \over {\lambda_+}} ={ {2(2\sqrt 5-3)} \over
{\sqrt{13} -1}}=1.13000...,
\end{equation} 
and the 
exponent $\tau$ of the scaling form (\ref{scaling}) for the
distribution of cluster sizes :
\begin{equation} 
\tau=1+{3 \over
\varphi} =1+{3 \over \sqrt{5}} =2.34164...,
\end{equation} 
in agreement with our numerical estimates (\ref{nbeta}), (\ref{numegamma}), (\ref{numetau}).

In this Letter, we have given given a percolation picture of the 
random spin-1 chain. Topological long-range order characteristic of 
the Haldane phase is due to the existence of a single cluster of 
spins related in a VBS manner whose size diverges in the 
thermodynamic limit.
The transition to the random singlet phase 
with increasing disorder
leads to the vanishing of 
the string order parameter with a exponent $2\beta =1.17$ while the 
percolation susceptibility diverges with an exponent $\gamma =1.13$.
These exponents are exact because the RG procedure is asymptotically 
exact\cite{Fisher}.

\acknowledgments

We would like to thank S. M. Girvin and R. A. Hyman
for useful discussions and sending us Ref.\cite{PhD}.

%%%%%%%%%%%%%%%%%%%%%%%%%%%%%%%%%%%%%%%%%%%%%%%%%%%%%%%%%%%%%%%%%
%%%%%%%%%%%%%%%%%%%%%%%%%REFERENCES%%%%%%%%%%%%%%%%%%%%%%%%%%%%%%


\begin{references}
\bibitem{active}
The correct definition of the VBS clusters implies that
the variable $\mu$ differs from the definition of active spins
introduced in \cite{HY}. Our scaling exponent $\varphi$ differs from
the exponent $\phi=2$ of active spins. The exponent $2\beta$ is not
equal to the value $2/{\lambda_+}=1.53518...$ found in \cite{HY}.
\end{references}
\end{document}